\documentclass[sigconf]{acmart}

\AtBeginDocument{%
  \providecommand\BibTeX{{%
    \normalfont B\kern-0.5em{\scshape i\kern-0.25em b}\kern-0.8em\TeX}}}

\usepackage{booktabs}
\usepackage{hyperref}
\usepackage{multirow}

\copyrightyear{2020}
\acmYear{2020}
\setcopyright{acmcopyright}
\acmConference[SIGIR '20]{Proceedings of the 43rd International ACM SIGIR Conference on Research and Development in Information Retrieval}{July 25--30, 2020}{Virtual Event, China}
\acmBooktitle{Proceedings of the 43rd International ACM SIGIR Conference on Research and Development in Information Retrieval (SIGIR '20), July 25--30, 2020, Virtual Event, China}
\acmPrice{15.00}
\acmDOI{10.1145/3397271.3401306}
\acmISBN{978-1-4503-8016-4/20/07}

\begin{document}
\fancyhead{}

\title{How UMass-FSD Inadvertently Leverages Temporal Bias}

%
\author{Dominik Wurzer}
\affiliation{%
  \institution{School of Information Management}
  \streetaddress{jkl}
  \city{Wuhan University}
  \country{China}}
\email{wurzer.dominik@whu.edu.cn}

\author{Yumeng Qin}
\authornote{Corresponding Author}
\affiliation{
  \institution{School of Information Management}
  \streetaddress{jkl}
  \city{Wuhan University}
  \country{China}}
\email{yumeng.qin@whu.edu.cn}


%
\begin{abstract}
First Story Detection describes the task of identifying new events in a stream of documents. The UMass-FSD system is known for its strong performance in First Story Detection competitions. Recently, it has been frequently used as a high accuracy baseline in research publications. We are the first to discover that UMass-FSD inadvertently leverages temporal bias. Interestingly, the discovered bias contrasts previously known biases and performs significantly better. Our analysis reveals an increased contribution of temporally distant documents, resulting from an unusual way of handling incremental term statistics. We show that this form of temporal bias is also applicable to other well-known First Story Detection systems, where it improves the detection accuracy. To provide a more generalizable conclusion and demonstrate that the observed bias is not only an artefact of a particular implementation, we present a model that intentionally leverages a bias on temporal distance. Our model significantly improves the detection effectiveness of state-of-the-art First Story Detection systems.

\end{abstract}

%
%


%

%

%
\maketitle

\section{Introduction}
First Story Detection (FSD), also called New Event Detection, is a research task introduced as part of the Topic Detection and Tracking (TDT) initiative \cite{AllanTopicDetectionBook}. The goal of FSD is to identify the very first document in a stream to mention a new event. This task has direct applications in news agencies and finance. The TDT initiative held several competitions during which the UMass-FSD \cite{umass} system was recognized for its strong performance in detection effectiveness \cite{fiscus2004results,petrovic2010streaming}. In recent years UMass-FSD has been actively used as a high accuracy baseline by state-of-the-art FSD systems \cite{wurzer2015twitter-scale,petrovic2010streaming,petrovic2012paraphrases,qin2017counteracting}, which try to scale to high volume streams while retaining a level of accuracy comparable to UMass-FSD. In this paper, we investigate the novelty computation algorithm of UMass and discover for the first time that it applies a new form of temporal bias in the decision making process. We show that this new form of bias towards the temporal is also applicable to modern FSD systems. In addition, we develop a new bias model that intentionally leverages the idea of bias on temporal distance. When learning optimal weights, 
\\\\\textbf{Contributions:}\begin{itemize}
\item \textbf{Discovering Temporal Bias in UMass-FSD}\\We are the first to report about a temporal bias in the UMass-FSD system.
\item  \textbf{a New Model for Temporal Bias on Distance}\\We present a new model that intentionally leverages a bias towards temporal distance and show how it significantly increases the effectiveness of state-of-the-art FSD systems.

\end{itemize}

\section{Related Work}
The Topic Detection and Tracking (TDT) initiative \cite{AllanTopicDetectionBook} defined First Story Detection (FSD) to be a streaming task. As documents arrive continuously one at a time, decisions on whether they speak about a new topic need to be made instantly and without prior knowledge about the detection targets or term statistics \cite{AllanTopicDetectionBook}. The most successful approach to FSD estimates the novelty of a new document arriving from the stream by its distance to the most similar previously received document \cite{wurzer2015twitter-scale,wurzer2018parameterizing}. This problem is also known as the single nearest neighbor (1-NN) task. The pilot study of TDT \cite{Allan98topicdetection} found that TF.IDF weighted Cosine Similarities work best when computing novelty in streaming settings like FSD. Traditionally, the Inverse Document Frequency (Equation \ref{eq:idf}) of a term ($t$) is computed by the ratio of the collection size ($|C|$) to the Document Frequency ($|\{d:t \in d, d \in C \}|$), i.e. the number of documents containing term $t$ \cite{wong1985generalized}. In streaming settings, the collection size and Document Frequencies change on the arrival of each new document. This requires a re-computation of all affected Inverse Document Frequency (IDF) statistics, also referred to as incremental IDF \cite{kannan2018real,petrovic2010streaming}. UMass-FSD \cite{umass}, a system known for its high effectiveness \cite{fiscus2004results,petrovic2010streaming,wurzer2018parameterizing}, estimates a document's novelty based on the distances to previously received documents, measured by incrementally weighted TF.IDF Cosine Similarities \cite{umassBounds}. In this paper, we show for the first time that UMass-FSD's unique incremental IDF update procedure can cause a temporal bias that significantly increases detection accuracy in FSD.\\\\
Temporal bias is not a new concept in FSD. Systems, like CMU-FSD \cite{CmuFSD}, reported a slight increase in effectiveness when prioritizing documents with high temporal proximity. Instead of computing novelty with respect to all previous documents, CMU-FSD retains a fixed-length window covering the most recent documents. Excluding older documents from the detection process through a sliding window implies a bias on recency. CMU-FSD further biases the novelty computation of new documents by decreasing the similarity to temporally distant documents.  
\vspace{-1.5mm}
\begin{equation}
\label{eq:CMU-FSD}
novelty(d_n) = 1 - \displaystyle\max_{d_i \in window}{\{cosSim(d_n,d_i) * \frac{i}{|window|}\}}
\end{equation}
Equation 1 shows that CMU-FSD  normalizes the Cosine Similarity between a new ($d_n$) and a previous document ($d_i$) by the fraction of the previous document's position ($i$) within the window and the window size ($|window|$) \cite{CmuFSD}. The fixed window size substantially reduces the search space, which increases efficiency. Additionally, a slightly increased degree of effectiveness was measured when applying Equation \ref{eq:CMU-FSD} to CMU-FSD on the official TDT data sets \cite{CmuFSD}. Modern FSD systems, like LSH-FSD \cite{petrovic2010streaming}, apply a similar but more relaxed form of bias towards temporal proximity. LSH-FSD \cite{petrovic2010streaming} was the first system to demonstrate that FSD is applicable to high volume social media streams, like Twitter\footnote{ \url{https://www.twitter.com/}}. LSH-FSD also applies a sliding window spanning the $n$ most recent documents. Whenever the window does not contain a sufficiently close document, LSH-FSD backs-off to an approximate nearest neighbor search (using Locality Sensitive Hashing -  LSH) covering all previously encountered documents. The original publication on LSH-FSD \cite{petrovic2010streaming} refers to this step as ``Variance Reduction Strategy'', which prioritizes recent documents over temporally distant ones for the novelty computation process in FSD. The motivation for biasing novelty detection towards recency arises from the idea that events emerge, grow, and subsequently fade away. Additionally, it was shown that documents belonging to an event tend to appear in ``clumps" \cite{AllanTopicDetectionBook}. Previous attempts \cite{CmuFSD,petrovic2010streaming} to temporally bias FSD are based on the idea that documents with increasing temporal proximity should have progressively less influence on current decision making. This strategy, although empirically proven successful, contradicts the definition of detection target by TDT. According to the official (TDT) specification of FSD \cite{AllanTopicDetectionBook}, detection targets are documents that are sufficiently different from \textit{all} previously seen documents.\\\\In this paper, we present a new form of temporal bias - geared towards the temporal distance, which can significantly improve the effectiveness of state-of-the-art FSD systems. By intuition, a bias on temporal distance decreases a document's novelty, if it is related to older topics. Interestingly, we discovered this form of temporal bias while studying an anomaly in the original UMass-FSD system, which is not known to apply temporal biases.
\begin{equation}
\label{eq:cosineSim}
\includegraphics[width=0.6\linewidth]{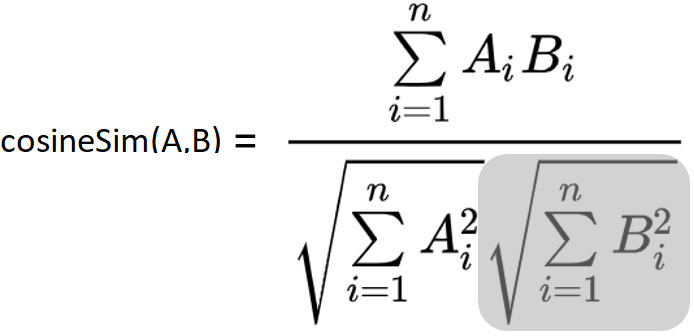}
\end{equation}

\begin{equation}
\label{eq:idf}
idf(t) = log(\frac{|C|}{|\{d:t\in d,d\in C \}|})
\end{equation}

\section{Temporal Bias through Incremental Term Statistics}
As is usual for streaming applications, UMass-FSD applies incremental term statistics and re-computes the IDF components on the arrival of each new document. None of the publications describing the UMass-FSD system reported about harnessing any form of temporal bias. According to the original publication \cite{umassBounds}, UMass-FSD solely bases its decisions on the distance to the closest previously encountered document, determined by the Cosine Similarity. \\\\Incremental term statistics are updated on the arrival of each new document. This requires a subsequent re-computation of previous document vector lengths, which act as a length normalization factor in the Cosine Similarity function (Equation \ref{eq:cosineSim}). We found that UMass-FSD accurately updates the term statistics, but omits the re-computation of the vector lengths for previously encountered documents (the highlighted grey area in Equation \ref{eq:cosineSim}). The UMass-FSD system descriptions do not mention the omission of the length update. We presume that it is skipped to increase the runtime efficiency - unaware of the impact it has on the detection accuracy. The following sections investigate the consequences of skipping vector length updates when applying incremental term statistics to a large number of documents.

\subsection{Incremental IDF Over Time}
UMass-FSD scales its document vectors by TF.IDF, a product of Term Frequency (TF) and Inverse Document Frequency (IDF). The IDF component (Equation \ref{eq:idf}) consists of the logarithmic ratio between the collection size ($|C|$ i.e., the number of previously received documents) and document frequency (i.e., the number of previously received documents containing term $t$). In order to understand the impact of neglecting document length updates, we first focus on analyzing the behavior of IDF components over time. Figure 1 shows the average IDF value for 1 million chronologically sorted English tweets. The graph illustrates that the average IDF value increases continuously over time. The steady increase occurs because the number of documents rises faster than the document frequencies, as most terms do not appear in all documents.
\begin{figure}[h]
\label{fig:idfOverTime}
  \centering
  \includegraphics[width=0.75\linewidth]{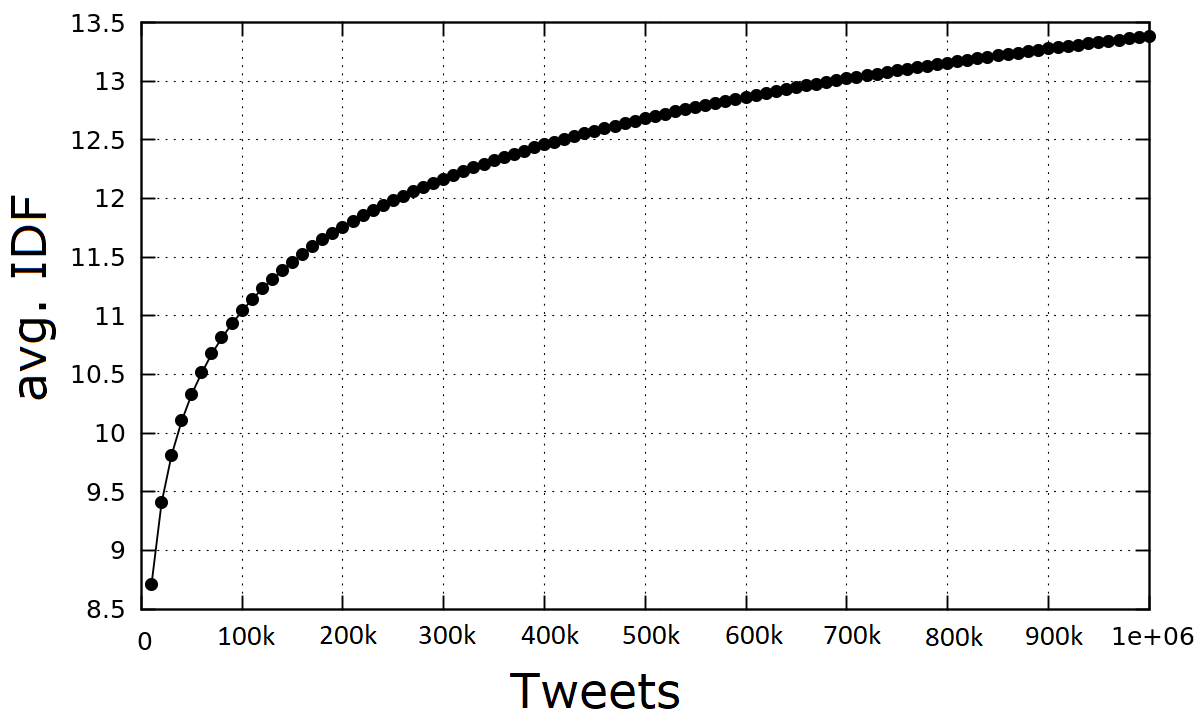}
  \caption{Average Inverse Document Frequency (IDF) according to Equation \ref{eq:idf} over 1 million chronologically (time-stamp) sorted tweet.}
\end{figure}
\subsection{Analyzing the Effect of Omitting Length Updates}
Continuously rising IDF values increase the magnitudes of the term vectors over time. Usually, this does not alter the FSD novelty estimation procedure, as document vector magnitudes are re-computed when the term statistics are updated. UMass-FSD omits the vector length updates. Instead, it relies on constant vector lengths defined by the term statistics at the time of a document's arrival. Equation \ref{eq:cosineSim} shows the Cosine Similarity between two documents and highlights the part that uses outdated term statistics (the vector length of the previous document B) in grey. We already ascertained that the average IDF and document vector magnitudes increase over time. The Cosine Similarity \\(Equation \ref{eq:cosineSim}) normalizes the dot product between two vectors by their vector lengths. Consequently, skipping the length updates decreases the normalization factor for older documents, which increases their similarity scores. UMass-FSD estimates a document's degree of novelty by the distance to its most similar previously received document. We therefore conclude that the novelty computation of UMass-FSD is biased towards temporarily distant documents. To put it concisely, UMass-FSD decreases the novelty of a document if it is \textit{relevant} to an older topic. The following section explores the impact of UMass's temporal bias on its effectiveness.



\section{Experiments}
In this section, we explore the impact of temporal bias on the effectiveness of First Story Detection. The reproducibility of our experiments is ensured by applying original FSD systems with default settings and evaluate them on publicly available data sets, using the standard TDT \cite{AllanTopicDetectionBook} evaluation metric and the official TDT evaluation script with default settings. We denote ``temporal bias'' by \textit{TB} to increase readability\\\\
\textbf{Data Set}\\
We explore the impact of TB on three public FSD research data sets: cross-twitter\footnote{The Cross Project is a joint venture between the University of Edinburgh and the University of Glasgow, \url{http://demeter.inf.ed.ac.uk/cross}}, a modern Twitter-based data set and the original TDT1 and TDT5 newswire data sets \footnote{TDT1 and TDT5 by Linguistic Data Consortium, NIST \url{https://catalog.ldc.upenn.edu/LDC2006T18}}. TDT1 consists of 25 topics and 15,863 documents and TDT5 consists of 126 topics and 278,108 documents. Cross-Twitter consists of 27 topics and comes in different corpora sizes. In this publication, we make use of cross-twitter-1.5mio, which consists of 1,500,000 tweets ordered by their publication time-stamp. The three data sets are frequently used in recent FSD publications \cite{petrovic2010streaming,wurzer2018parameterizing,wurzer2015twitter-scale,petrovic2012paraphrases,qin2017counteracting,moran2016enhancing}.\\\\
\textbf{Evaluation Metric}\\Following the official TDT guideline \cite{AllanTopicDetectionBook}, we evaluate the detection effectiveness by the normalized Topic Weighted Minimum Detection Cost ($C_{min}$). The detection cost $C_{min}$ linearly combines miss and false alarm probabilities to provide a single value metric for comparing different systems \cite{AllanTopicDetectionBook,umassBounds}. As is usual in the evaluation of FSD in the Topic Detection and Tracking (TDT) program \cite{wurzer2018parameterizing,CmuFSD,Allan98topicdetection}, we apply Skip Evaluation. Skip Evaluation increases the number of detection targets by iterating over the topics and replacing (\textit{skips}) all detection targets by their first follow-up documents. To prevent small scale topics from vanishing, we limited Skip Evaluation to 9 rounds for cross-twitter and 3 rounds for TDT1 and TDT5. \vspace{-2mm}
\subsection{The Effect of Temporal Bias on FSD effectiveness}
To assess the impact of temporal bias (TB) on FSD effectiveness, we compare the original UMass-FSD system (denoted by ``\textit{TB on distance}'') to UMass-FSD with correct term statistic updates (denoted by ``\textit{no TB}"). Furthermore, we add a TB on recency (Equation \ref{eq:CMU-FSD}), denoted by ``TB on recency''. Table 1 shows a diverging result for TB on recency. On small scale data sets, like TDT1, the detection cost is decreased (lower is better) by 3.27\%. The positive impact diminishes on the medium-sized TDT5 data set and becomes negative when applied to the large scale cross-twitter data set. When applied to large collection sizes, the fixed-sized window trades effectiveness against efficiency. Small window sizes are likely to miss the true nearest neighbor, whereas the bias overpowers the Cosine Similarity in larger windows. On the contrary, TB on distance significantly $(p<0.05)$ decreases the detection cost ($C_{min}$) on TDT5 and cross-twitter. Our analysis revealed that the positive effect of TB on distance increases when a topic's lifespan covers a large volume of documents. When applied to small scale data sets, like TDT1, the positive effect diminishes following neglectable changes in incremental term statistics. A side effect of biasing by omitting the vector length updates is an increase in efficiency by 66\% on cross-twitter for UMass-FSD.
\\\\The TB on distance, found in UMass-FSD, is also applicable to modern FSD systems. LSH-FSD narrows the search field through randomized approximation and exhaustively search the resulting candidate set by incrementally weighted TF.IDF Cosine Similarity. This allows biasing LSH-FSD towards temporally distant documents by omitting the vector length updates. Table 2 shows that TB on distance significantly $(p<0.05)$ improves the effectiveness of LSH-FSD on TDT5 and cross-twitter. Interestingly, we measured a slightly lower decrease in detection cost for LSH-FSD in comparison with UMass-FSD. Instead of identifying the true nearest neighbor, LSH-FSD approximates it and stops the search once a \textit{sufficiently} close (variance reduction)\cite{petrovic2010streaming} document is found. The combination of the approximated search and ``Variance Reduction Strategy'', limits the impact of the temporal bias on detection effectiveness. When applied to small data sets (TDT1), we found that LSH-FSD omits its approximation strategy and relies on exhaustive search - like UMass-FSD. This explains the comparable performance of the two systems on TDT1.\vspace{-2mm}
\subsection{Optimizing the Temporal Bias}
The previous section revealed that prioritizing relevant and temporally distant documents, can reduce the detection cost in FSD systems. In this section, we create our own bias on temporal distance (Equation 4). 
\begin{equation}
\mbox{bias}(d_n,d_i) = \begin{Bmatrix} 1 + \delta  * log( n - i ) & : cosineSim(d_n, d_i)   \eqslantgtr   \gamma  \\1  &  : cosineSim(d_n, d_i)   <   \gamma \end{Bmatrix} 
\end{equation}
We base the bias on the logarithmic difference between a document's ($d_n$) position ($n$) to a previously encountered document ($d_i$). \\The subscript ($i:i  \in  \{ 1 ... n \}$) indicates the previous document's position within the stream. We optimize the weights ($\delta  = 0.036,\gamma = 0.61$) on a training data set\footnote{Wurzer et, al. (2018) Parameterizing Kterm Hashing - consists of 15 topics + 10 rounds skip evaluation} by an SVM\cite{Boser92atraining}, using a radial basis function kernel with a convergence tolerance of 0.01 and class weights \citep{wurzer2018parameterizing} to address the class imbalance. The bias becomes effective when the Cosine Similarity exceeds the threshold parameter $\gamma$, which is $\gamma$ = 0.61 on our training data set. The threshold parameter limits the bias to ``relevant'' documents instead of all older documents. The novelty score of a new document ($d_n$) is computed by: $Novelty(d_n) = 1 - \max  \{ cosineSim(d_n, d_i) * bias (d_n, d_i) \}$. For example, in Cross-Twitter, document $d_{286}$ reports about a new event. A follow-up document ($d_{10,067}$ ) reports about the same event 9,781 documents later.  The novelty score for document $d_{10,067}$ resulting from the default cosine similarity is 0.3871. Based on the temporal distance, Equation 4 inflates the cosine similarity and reduces the novelty score of the follow-up document ($d_{10,067}$) to 0.365. Table 1 and Table 2 show the impact of the optimized temporal bias (denoted ``TB optimized") on the detection effectiveness for UMass-FSD and LSH-FSD. Both systems reach their highest effectiveness (lowest detection cost) on TDT5 and cross-twitter when applying our optimized TB (Equation 4). The difference reaches statistical significance $(p<0.05)$ when applied to large data sets. The positive effect diminishes on small-scale data sets (TDT1) due to lower document volumes during the event time-spans.





\begin{table}[]
\begin{tabular}{|c|c|c|c|c|}
\hline
\multicolumn{5}{|c|}{UMass-FSD}\\ \hline
\multirow{3}{*}{\textbf{data set}} 	& \multirow{2}{*}{\textbf{no TB}} & \textbf{TB on} & \textbf{TB on} & \textbf{TB}   \\ 
									& & \textbf{recency}  & \textbf{distance**} & \textbf{optimized}   \\ 
									& ($C_{min}$/Dif.) &  ($C_{min}$/Dif.) &  ($C_{min}$/Dif.) & ($C_{min}$/Dif.) \\ \hline

\multirow{2}{*}{TDT1} 	& 0.642 	& \textbf{0.621}		& 	0.641		& 0.640 					\\ 
 						& -			& \textbf{(-3.27\%)}  	& 	(-0.16\%)	& (-0.31\%)				\\ \hline 
 	
\multirow{2}{*}{TDT5}	& 0.756 	& 0.751 	   	& 0.697*		& \textbf{0.688*}		\\ 
						& - 		& (-0.66\%) 	& (-7.80\%)		& (\textbf{-7.99\%}) 	\\ \hline 
	
cross					& 0.872 	&  	   0.936	& 0.798*		& \textbf{0.787*}		\\ 
twitter					& - 		& (+7.3\%)	   	& (-8.47\%)		& (\textbf{-9.75\%})	\\ \hline  
\end{tabular}
 \caption{Comparing the impact of temporal bias on UMass-FSD's detection cost ($C_{min}$: lower is better) for 3 data sets. Temporal bias is denoted by \textit{TB}. Asterisk (*) indicates statistical significance, (**) indicates the original system.}
 \vspace{-5mm}
\end{table}

\begin{table}[]
\begin{tabular}{|c|c|c|c|c|}
\hline
\multicolumn{5}{|c|}{LSH-FSD}\\ \hline
\multirow{3}{*}{\textbf{data set}} 	& \multirow{2}{*}{\textbf{no TB**}} & \textbf{TB on} & \textbf{TB on} & \textbf{TB}   \\ 
									& &  \textbf{recency} & \textbf{distance} & \textbf{optimized}   \\ 
									& ($C_{min}$/Dif.) &  ($C_{min}$/Dif.) &  ($C_{min}$/Dif.) & ($C_{min}$/Dif.) \\ \hline

\multirow{2}{*}{TDT1}	& 0.679 	& \textbf{0.665}  		& 0.677			& 0.675 				\\ 
 						& -			& \textbf{(-2.06\%)}  	& (-0.29\%)		& (-0.59\%) 			\\ \hline 
 	
\multirow{2}{*}{TDT5} 	& 0.762 	& 0.758 		& 0.714* 	&	\textbf{0.706*} \\ 
						& - 		& (-0.53\%) 	& (-6.30\%)	& (\textbf{-7.40\%})\\ \hline  
	
cross					& 0.906 	& 0.942 		& 0.837*	& \textbf{0.831*}	\\ 
twitter					& - 		& (+3.97\%)	   	& (-7.62\%)	& (\textbf{-8.28\%})\\ \hline 
\end{tabular}
 \caption{Comparing the impact of temporal bias on LSH-FSD's detection cost ($C_{min}$: lower is better) for 3 data sets. Temporal bias is denoted by \textit{TB}. Asterisk (*) indicates statistical significance, (**) indicates the original system.}
 \vspace{-5mm}
\end{table}

\section{Conclusion}
This paper described how UMass-FDS inadvertently leverages a bias on temporal distance. The novelty scores of new documents are lowered, if they are relevant to older and previously known topics. We showed that the bias results from an implementation decision not to update the document vector lengths corresponding to older documents. This places higher weights on older topic and increases their contribution to the decision-making process. Our analysis showed that the discovered bias is also applicable to other well-known FSD systems, where it improves the detection effectiveness on large-scale data sets. Additionally, we presented a new bias model that intentionally leverages the idea of bias towards the temporal distance. Our experiments demonstrated that our model significantly increases the detection effectiveness of state-of-the-art FSD systems.\vspace{2mm}


%
\begin{acks}
We thank Dr. Victor Lavrenko for providing the source code and guidance for the original UMass-FSD and LSH-FSD systems. We also thank Dean Qing Fang for providing the computing power necessary to carrying out the experiments.

\end{acks}

%
\bibliographystyle{ACM-Reference-Format}
\bibliography{bibi}

\end{document}